\documentclass[conference]{IEEEtran}
\usepackage{cite}
\usepackage{amsmath,amssymb,amsfonts}
\usepackage{textcomp}
\usepackage{xcolor}
\usepackage{graphicx}  
\usepackage{multirow}
\usepackage{algorithmic}
\usepackage{array}
\def\BibTeX{{\rm B\kern-.05em{\sc i\kern-.025em b}\kern-.08em
    T\kern-.1667em\lower.7ex\hbox{E}\kern-.125emX}}
 \begin{document}
%
\title{Improving the Robustness to Data Inconsistency between Training and Testing for Code Completion by Hierarchical Language Model}
\author{
\IEEEauthorblockN{Yixiao Yang}
\IEEEauthorblockA{\textit{College of Information Engineering, Capital Normal University} \\
Beijing, China \\
yangyixiaofirst@outlook.com}
}

\maketitle
\begin{abstract}
Language models are state-of-art models which have been widely used in many applications in the field of natural language processing. 
In the field of software engineering, applying language models to the token sequence of source code is the state-of-art approach to build a code recommendation system. 
The syntax tree of source code has hierarchical structures. Ignoring the characteristics of tree structures decreases the model performance according to the experiment. 
Current LSTM model handles sequential data. The performance of LSTM model will decrease sharply if the noise unseen data is distributed everywhere in the test suite. As code has free naming conventions, it is common for the model trained on one project to encounter many unknown words on another project. If we set many unseen words as UNK just like the solution in natural language processing,  the number of UNK will be even much greater than the sum of the most frequently appeared words. 
In an extreme case, just predicting UNK at everywhere may achieve very high prediction accuracy. 
Thus, such solution cannot reflect the true performance of a model when encountering noise unseen data. 
In this paper, we only mark a small number of rare words as UNK and show the prediction performance of models under in-project and cross-project evaluation. 
We propose a novel Hierarchical Language Model (HLM) to improve the robustness of LSTM model to gain the capacity about dealing with the inconsistency of data distribution between training and testing. The newly proposed HLM takes the hierarchical structure of code tree into consideration to predict code. 
HLM uses BiLSTM to generate embedding for sub-trees according to hierarchies and collects the embedding of sub-trees in context to predict next code. 
The experiments on inner-project and cross-project data sets indicate that the newly proposed Hierarchical Language Model (HLM) 
performs better than the state-of-art LSTM model in dealing with the data inconsistency between training and testing and achieves averagely 11.2\% improvement in prediction accuracy. 
\end{abstract}


\section{Introduction}
Language models have become the cornerstone of other advanced technologies. 
Many applications such as machine translation \cite{MT}, audio recognition \cite{AR} and text classification \cite{TC} adopt language models to improve the model performance. 
In recent period, the field of software engineering has adopted language models to improve the performance of many tasks such as code smell detection \cite{CodeSmell} and  software bug detection \cite{BugDetectNGram}. 
To build a code recommendation system, the source code is parsed into token sequence and the language model is applied to learn that token sequence to help recommend code snippets or code fragments. 
Current state-of-art methods include statistical language model or neural network based language model (also named as deep language model). 
The existing works have contributed to the solution of of code completion and helped software engineers improve the efficiency of developing software. 

However, in sequential language models, the structures of source code is ignored. Ignoring the valuable structure information will degrade the performance of the model. 
Besides, the sequential language model such as LSTM is weak in handling the too long token sequence of code which contains unseen code snippets. Such problem is named as \emph{exponential bias}. If some parts in testing phase are unseen in training phase, the discrepancies will occur. The discrepancies may gather and explode along a long prediction phase. 
The standard LSTM cannot deal with the exponential bias problem according to the experiments. One possible solution is to set many words as UNK to avoid unseen data. In this paper, we address the problem in model level and no longer need to set many UNK words in training set. The traditional techniques only take the fixed-length previous tokens in the sequence into consideration. In this paper, we try to predict every token in a function to examine the performance. 

For long code token sequence, the LSTM may fail to capture the characteristics of the whole token sequence. 
The idea which divides the long previous tokens into segments and predict next code based on the generated segments is proposed. 
As the abstract syntax tree (AST) of source code is in a hierarchical structure. The sub-trees in different hierarchies can be taken as different segments. 
In another word, the tokens in a sub-tree can be grouped into one segment and the next code can be predicted based on sub-trees rather than tokens. 
This can help model handle long code context. 
Nodes at different hierarchies in AST have different correlations with each other. 
Capturing the hierarchical structure of source code could help identify the relationship of nodes on the corresponding AST. 

For the problem of predicting node on tree, based on the hierarchical structure of code, we propose the
Hierarchical Language Model (HLM). For each hierarchy of the code tree, we use the standard LSTM to 
accumulate the information of each sub-tree in that hierarchy (also accumulate the information of each sub-tree in higher hierarchies) to improve the prediction performance. 
The information of each sub-tree is generated based on BiLSTM. 
To generate the embedding for each sub-tree, existing models such as Tree Neural Network (TreeNN) \cite{TreeNN2} could be applied. 
However, there are problems in existing tree encoding models. 
The strong abstract ability of existing tree models has brought problems to training. 
LSTM is non-convex. The LSTM model which integrates tree encoding models together is more complex.
When training with the summation of so many complex loss functions, the model is difficult to jump out the local optimal solution. 
The recent tree encoding work \cite{DBLP:conf/icse/ZhangWZ0WL19} is adopted to use BiLSTM to encode the tree to make the training of the model more smooth. 
The results of inner-project and cross-project experiments indicate that Hierarchical Language Model (HLM) 
outperforms state-of-art models in handling data inconsistency. 
The HLM model can be taken as a general framework to generate data in tree structure. Theoretically speaking, The HLM model could be applied to any language which can be parsed into an abstract syntax tree (AST). 
The main contributions of this paper include: 
\begin{itemize}
\item A model is proposed to predict code in the  hierarchical structure of AST. 
\item The tree encoding mechanism is designed and applied on sub-trees. The encoding of sub-trees is applied to help predict code. 
\item Both inner-project and cross-project evaluation are conducted to compare the performance of models. 
\end{itemize}

\section{Related Work}
\subsubsection{Statistical models for code completion} 
The statistical language model is based on the statistical information of source code. The information includes the statistical patterns of code. 
The pioneer work \cite{DBLP:conf/icse/HindleBSGD12} using language model is based on the statistical n-gram model. 
The naturalness and predictability of source code are verified by counting the number of n-grams in the program. 
SLAMC \cite{DBLP:conf/sigsoft/NguyenNNN13} assigns topics to each token and uses n-gram model to predict next code token based on tokens in the context and the corresponding topics of tokens in the context. 
A large-scale investigation \cite{DBLP:conf/msr/AllamanisS13a} of n-gram model on large code corpus are conducted and a visualization tool is provided. 
Cacheca \cite{Tu2014On} confirms the localness of source code and proposes a cache model 
to improve the code suggestion performance. 
The patterns of common API call sequences together with the associated parameter objects are captured by per-object n-grams \cite{DBLP:conf/pldi/RaychevVY14}. 
The Naive-Bayes model was applied on graph \cite{DBLP:conf/icse/NguyenN15} to suggest API usages. 
Decision tree together with n-gram model \cite{DBLP:conf/oopsla/RaychevBV16} was applied to solve the problem of code completion. 
The code was modelled in the form of DSL \cite{poplRaychevBVK16}. Based on DSL, the model was trained in the way that the model keeps sampling and validating the recommended code until the right code was suggested. 
For statement level code completion, the work \cite{DBLP:conf/kbse/YangJ0SGL17} uses IR and fuzzy search algorithm to search for similar context to handle the unseen data to improve the standard n-gram model. 

\subsubsection{Deep learning models for code completion} 
The pioneer work \cite{White2015Toward} using deep language model to solve the problem of code completion is based on RNN model. 
Long Short Term Memory (LSTM) network is a kind of recurrent neural network which introduces the gate mechanism to capture longer dependencies than RNN model. The LSTM model was applied to solve the problem of code completion \cite{dam2016deep} \cite{FSE17}  to achieve higher accuracy. 
Attention network \cite{DBLP:conf/ijcai/LiWLK18} is applied to LSTM model to further improve the ability of capturing the characteristics of the context. 
Pointer network  \cite{DBLP:conf/ijcai/LiWLK18} or graph network \cite{DBLP:conf/seke/YangX19} is adopted to predict the unseen data. The main difference between the two work is that the work in \cite{DBLP:conf/seke/YangX19} adopts different switch algorithm and separates the parameters between language model and repetition learning model. This makes the prediction effect of unknown data more obvious. 
The work \cite{DBLP:conf/seke/YangX19a} uses the tree language model with the tree encoding of context. The tree encoding technology is the 2D-LSTM model. 
To improve the robustness of the language model. Sequence-GAN \cite{yu2017seqgan} 
generates both the right and wrong examples to improve the robustness of the model. Different sampling schedules \cite{Bengio2015Scheduled} are also applied to improve the robustness of the model. 
\subsubsection{Models for code synthesis}
Based on the hint described in natural languages or other forms, code synthesis is to generate a whole code snippet. 
Many models such as Seq2Seq \cite{iyer2018mapping}, Seq2Tree \cite{yin2017syntactic} and Tree2Tree \cite{drissi2018program} models were proposed for the problem of code synthesis. 
To synthesize the API sequences based on natural languages, Seq2Seq model \cite{Nguyen2016T2API,APILearn} was applied. 
The neural program translation needs to translate one program to another program. The program is in the form of abstract syntax tree(AST) and the Tree2Tree \cite{chen2018tree} model translates one tree to another tree. 
\subsubsection{Models for code classification} 
Each program is assigned an embedding matrix \cite{Piech2015Learning} which is used to encode the pre-conditions and reconstruct the post-conditions for a program. 
To generate the name of a function, the convolution based attention model \cite{allamanis2016convolutional} is applied. 
The Tree Based Convolution Neural Network (TBCNN) \cite{TreeBasedCNN} applies the general convolution mechanism to the syntax tree of code to judge what type of program it is. 
To efficiently search for computationally efﬁcient identities, 
TreeNN \cite{zaremba2014learning} was adopted in producing representations of  homogeneous and polynomial expressions. 
To categorize expressions according to the semantics, based on TreeNN, EQNET \cite{EQNET} additionally adopted an extra abstraction and reconstruction layer to help model capture the semantics of the expressions. 
Code is organized in statements and BiLSTM model is applied to the embedding of each statement \cite{DBLP:conf/icse/ZhangWZ0WL19} to generate the representation for code snippet to help classify the code snippet. 
\subsubsection{Hierarchical Language Model for source code} 
As far as we know, 
HLM is the first one to use the BiLSTM embedding of sub-trees on the hierarchical structure of code to explicitly handle the data inconsistency between training and testing. 
The experiments indicate that the newly proposed model indeed has some special properties. 

\section{Proposed Method}
\subsection{Preliminary}
\begin{figure}[htbp]
\centering
\includegraphics[width=0.88\linewidth]{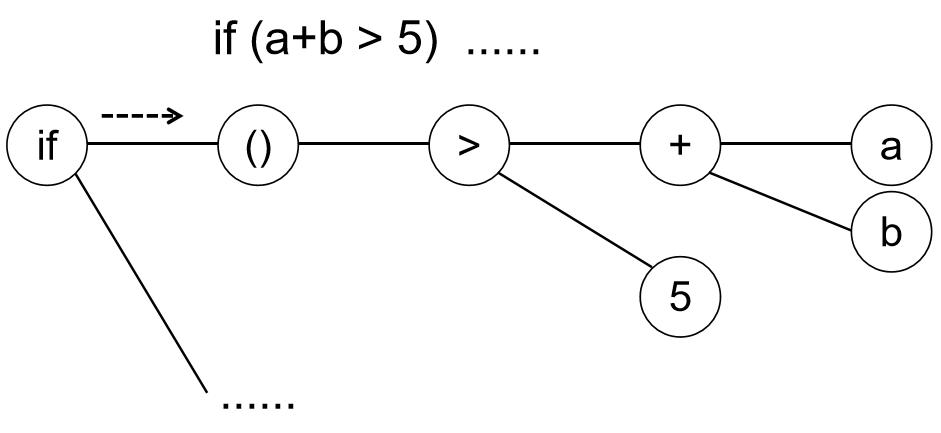}
\caption{An Example of AST}
\label{fig:ast_translation_example}
\end{figure}

\noindent {\bfseries AST(Abstract Syntax Tree):} The source code and its corresponding AST are illustrated in Figure \ref{fig:ast_translation_example}. Each node in AST has a content (also called token). For instance, a node with content (token) "$+$" meaning the plus operation. In this figure, the root is the node with content $if$ and the leaf nodes are node $a$, node $b$ and node $5$. 
In this paper, the horizontally placed trees are conducive to explain the proposed ideas. 
\\
\noindent {\bfseries Code Completion on AST:}
For an AST, the aim of code completion is to complete each node in that AST one by one. As each node must be predicted one by one, the order in which each node is generated must be decided. We observe that when people write a function, the skeleton of that function (the declaration of the function) is often written first. Then writing statements in the body of the function one by one and so on. This order in which the code is written is consistent with the pre-order traversal of AST. Thus, when predicting nodes on an AST, we use pre-order traversal to traverse the tree to predict each encountered node. The HLM model is for predicting each node. This means that the HLM model traverses a tree in pre-order from the root to the leaves to predict each node. When using the pre-order traversal to traverse a tree, the already visited nodes are taken as context in the problem of code completion. The encoding model in HLM is to encode the complete trees or sub-trees existed in the context. Current models for generating encoding for a tree traverse that tree in post-order to generate representations from the leaves to the root. The encoding model in HLM inherits this setting meaning that Encoding model also uses the post-order traversal to traverse a tree to accumulate the information from the leaves to the root to generate the encoding for a tree. 
\\
\noindent {\bfseries Concepts on AST:}
The formal definition of some concepts of AST is described here. 
\begin{itemize}
\item Sibling nodes: In AST, the sibling nodes have  common parent. They are also considered to be at the same level or at the same hierarchy. For example, in Figure \ref{fig:ast_translation_example}, node $5$ and node $+$ are sibling nodes, node $5$ and node $+$ are considered to be at the same level or at the same hierarchy.
Node $+$ is the previous sibling of node $5$, node $5$ is the next sibling of node $+$. 
\item Descendent nodes: For node n, all nodes in the tree rooted at node n are referred to as descendants of node n.
\item Ancestor nodes: If node p is the descendent of node q, then node q is referred to as the ancestor of node p. 
\end{itemize}

\subsection{Hierarchical Language Model}
The decoding process of a language model is to accumulate the information of context to predict the next node. For node n in AST, the context of node n is defined as follows: In the pre-order traversal of a tree, the nodes traversed before traversing the node n make up the context of the node n. To fully use the context information of a node, the language model needs to collect the information of every node in the context of that node to help predict the following subsequent nodes. For traditional language model, to predict node n, every node of the context needs to be iterated and processed by LSTM. The order in which nodes are processed by LSTM is consistent with the order in which nodes are traversed. We define a concept: $decoding\ path$ to illustrate the order of the processing of nodes in the decoding procedure. The $decoding\ path$ is a node sequence in which every node is processed step by step. 

For the proposed Hierarchical Language Model (HLM), 
to predict node n at the specified position, HLM accumulates information based on $decoding\ path$ of node n. 
The decoding process in HLM walks through the $decoding\ path$ to accumulate the necessary information to predict the next nodes. 
Details will be shown in the following subsections. 
Here, to make it easier to explain the mechanism and have an intuitive comparison of HLM and the traditional language model, for predicting a single node, we use the corresponding $decoding\ path$ to show the details of the decoding mechanism. Actually, the whole process could be carried out in a continuous depth-first manner. 
\begin{figure}[htbp]
\centering
\includegraphics[width=0.92\linewidth]{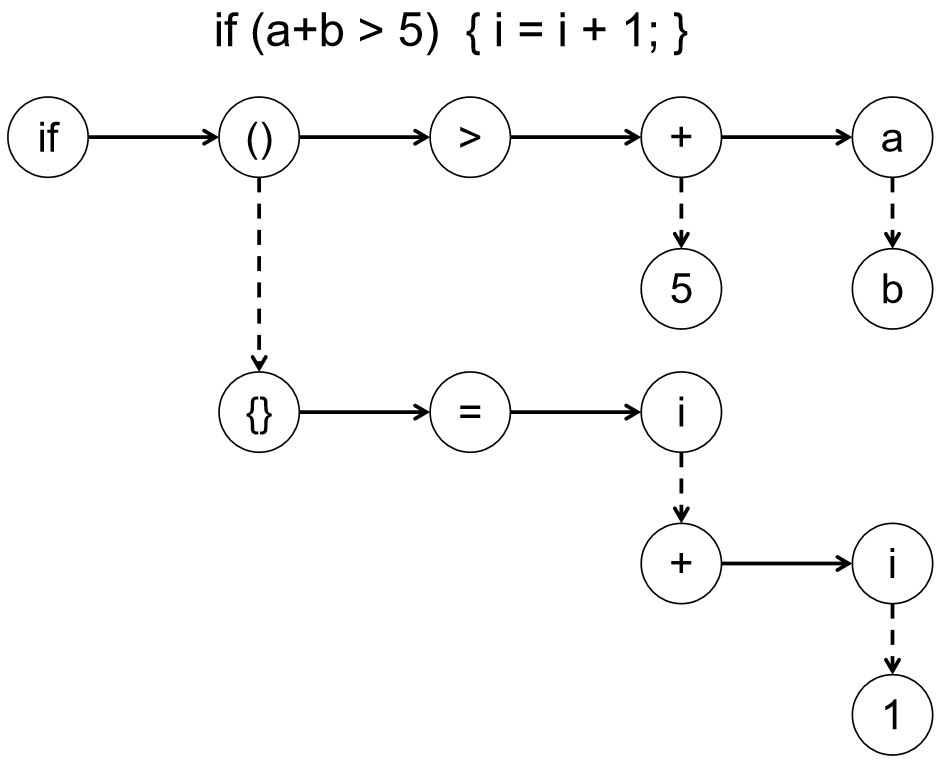}
\caption{Path from Root to Node}
\label{fig:transformed-ast}
\end{figure}

\noindent {\bfseries Decoding Path of HLM:} The $decoding\ path$ of HLM for node n is the transfer path from the root to node n. From a node, only the first child of that node or the next sibling of that node can be transferred to. Thus, the candidate transfer paths of the AST in Figure \ref{fig:ast_translation_example} are shown in Figure \ref{fig:transformed-ast}. In Figure \ref{fig:transformed-ast}, the solid arrow represents the transition to the first child, and the dotted arrow represents the transition to the next sibling node. 
Thus, under this definition, a Directed Acyclic Graph (DAG) has been generated and the transfer path from the root to each node has been uniquely determined. 

In details, from the root node of the tree, if node n is the descendent of the root node, to reach node n, we must transfer from the root node to the first child of the root node. After reaching new node, then, if node n is the descendent of the newly reached node, we must transfer to the first child of the newly reached node. Otherwise, we must transfer to the next sibling of the newly reached node. If we keep transferring in this way, we could finally reach the node n. The information is processed and accumulated when walking through the transfer paths. There are two kinds of transitions: transfer to first child and transfer to next sibling. 
For dealing with the transfer to the first child, traditional LSTM is adopted. 
For dealing with the transfer to the next sibling, 
LSTM together with hierarchical tree embedding is adopted. 

\begin{figure}[htbp]
\centering
\includegraphics[width=0.92\linewidth]{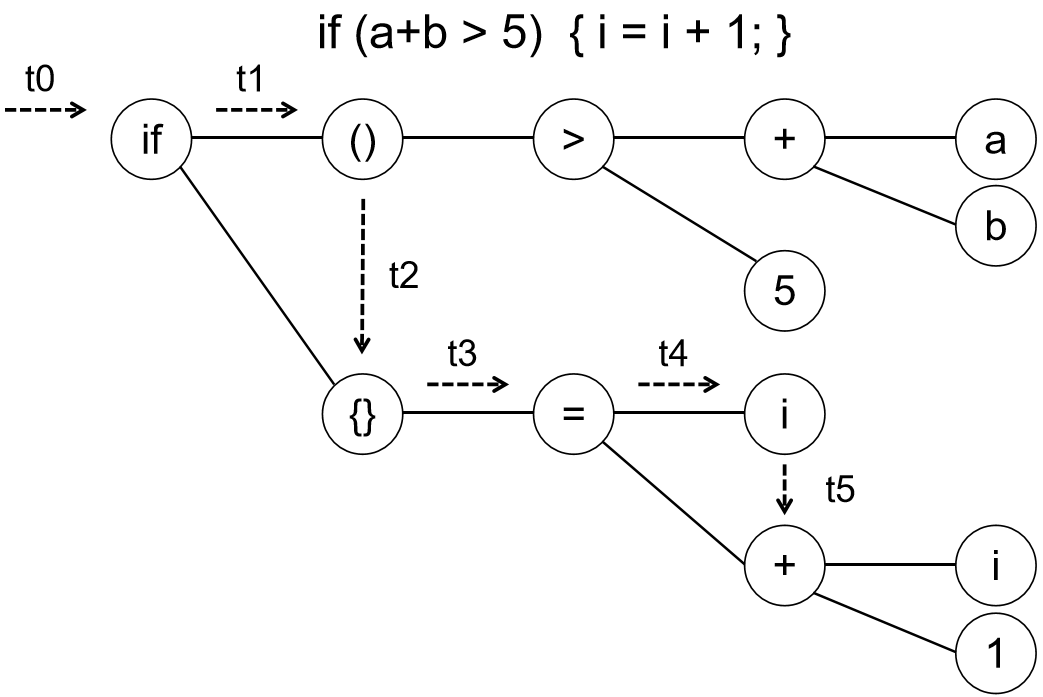}
\caption{Candidate paths of AST}
\label{fig:decoding-path}
\end{figure}

\noindent {\bfseries Transition on Decoding Path of HLM:}
As described above, the $decoding\ path$ consists of a sequence of transitions. 
In Figure \ref{fig:decoding-path}, the dotted arrows give an illustration about the path and the transitions from the root to node + (the second child of node =). 
Each transition between nodes on the path have been marked as $t_0$, $t_1$, $t_2$, ... $t_5$. 
The transition corresponds to the information flow. 
The information flowing on a transition represents the accumulated information of previous transitions before that transition. 
The information which flows on each transition has a fixed data format: (cell, h). The cell and h are two feature vectors of fixed length. Through learning from LSTM, two feature vectors: cell and h instead of one feature vector: h are used for representing the accumulated information of context. 
The symbols $cell_i$ and $h_i$ represent the information on transition $t_i$. 

\noindent  {\bfseries Decoding Procedure of HLM:} 
Based on the transitions of the $decoding\ path$ of node n, we start to iterate the transitions one by one to compute the decoding information for node n. At first, the information of transition $t_0$: $cell_0$, $h_0$ is set to a fixed default value. Then for each transition $t_i$, if $t_i$ is of kind: transfer to the first child, we use standard LSTM to compute the information for transition $t_i$. Assume the token embedding of the source node of the transition $t_i$ is referred to as $\tau_i$. The information of transition $t_i$ is computed by:
\begin{equation}
cell_i, h_i = LSTM(\tau_i, cell_{i-1}, h_{i-1})
\end{equation}
The computed $cell_i,\ h_i$ can be used to decode the target node of transition $t_i$. The $cell_i,\ h_i$ can also be used to compute the information of transition $t_{i+1}$. 
For the encountered transition $t_i$, if $t_i$ is of kind: transfer to the next sibling, we use a standard 2-dimensional LSTM (abbreviated as 2DLSTM) to compute the information for transition $t_i$. Assume the token embedding of the source node of the transition $t_i$ is referred to as $\tau_i$. The embedding of the descendants of the source node of transition $t_i$ is referred to as $cell_{des},  h_{des}$. The information of transition $t_i$ is computed by:
\begin{equation}
cell_i, h_i = 2DLSTM(\tau_i, cell_{des}, h_{des}, cell_{i-1}, h_{i-1})
\label{eq:2d-decode-logit}
\end{equation}
As mentioned above, the computed $cell_i,\ h_i$ for transition $t_i$ which transfers to next sibling can be used to decode the target node of transition $t_i$. The $cell_i,\ h_i$ can also be used to compute the information of transition $t_{i+1}$. Note that in this computation, all nodes in context before the transition $t_i$ takes place are actually taken into consideration. Although we do not walk through the descendants of the source node of transition $t_i$ in decoding procedure, but the information of the descendants of the source node of transition $t_i$ is computed through encoding techniques. To decode the token of target node of transition $t_i$ based on computed $cell_i,\ h_i$, we compute the probabilities of all tokens and the top-k tokens with highest probabilities are taken as the final results. The probabilities of all tokens are computed as:
\begin{equation}
probs = softmax(V h_i)
\label{eq:probs}
\end{equation}
In Equation \ref{eq:probs}, $V \in \mathbb{R}^{v\times d}$, $v$ is the size of vocabulary, $d$ is the length of the feature vector, the length of vector $h_i$ is also $d$. In Equation \ref{eq:probs}, $h_i \in \mathbb{R}^{d \times 1}$. 
The training procedure is to minimize the following loss function:
\begin{equation}
loss = -log(probs[oracle_i])
\label{eq:loss}
\end{equation}
The $oracle_i$ is the index of the token of target node of transition $t_i$ in the vocabulary table.

\begin{figure}[htbp]
\centering
\includegraphics[width=0.5\linewidth]{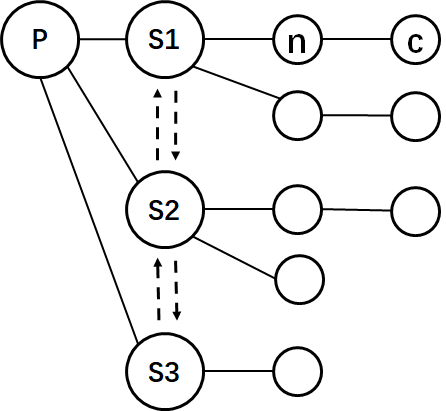}
\caption{Encoding of AST}
\label{fig:ast-encode}
\end{figure}

\noindent  {\bfseries Encoding of Descendants:} 
For descendants of a node, we design a hierarchical BiLSTM to encode all these nodes. 
For each sub-tree, unlike traditional encoding method such as TreeNN which generates only one vector $h^{tree}$ to represent the tree, we use two vectors $cell^{tree}$, $h^{tree}$ to represent the tree. We use post-order traversal to traverse the AST from leaves to the root to encode the tree. For leaf node, according to the token of that node, we use distinct trainable parameters as $cell^{tree}$ and $h^{tree}$ of that leaf node. As shown in Figure \ref{fig:ast-encode}, for non-leaf node n of which the number of children is 1 and the symbols $cell_c^{tree},\ h_c^{tree}$ represent the encoding of the sub-tree rooted at the only child c, we use the representation of tree rooted at the only child as the embedding of descendants of node n: $cell_{des} = cell_c^{tree};\ h_{des} = h_c^{tree}$. The encoding of tree rooted at node n is computed by: $cell_n^{tree}, h_n^{tree} = LSTM(\tau, cell_c^{tree}, h_c^{tree})$. The symbol $\tau$ refers to the embedding of the token of node n. 

As shown in Figure \ref{fig:ast-encode}, 
for non-leaf node P of which the number of children is 3 and the symbols $cell_{s1}^{tree},\ h_{s1}^{tree}$, $cell_{s2}^{tree},\ h_{s2}^{tree}$, $cell_{s3}^{tree},\ h_{s3}^{tree}$ represent the encoding of the sub-trees rooted at the children: $s1$, $s2$, $s3$, we use BiLSTM to iterate the children. For LSTM iterated from $s1$ to $s3$, the initial cell, h are $cell_{s1}^{tree},\ h_{s1}^{tree}$, then, the LSTM takes h of the encoding (cell, h) of the encountered sub-tree as the embedding of that sub-tree. Thus for s2 and s3, the LSTM handles $h_{s2}^{tree}$, $h_{s3}^{tree}$ step by step and finally outputs $cell_{f1}$ and $h_{f1}$. For another LSTM iterated from $s3$ to $s1$, the initial cell, h are $cell_{s3}^{tree},\ h_{s3}^{tree}$, then, the LSTM handles $h_{s2}^{tree}$, $h_{s1}^{tree}$ step by step and finally outputs $cell_{f2}$ and $h_{f2}$. The encoding of tree rooted at node P is computed by: $cell_P^{tree}, h_P^{tree} = 2DLSTM(\tau, cell_{f1}^{tree}, h_{f1}^{tree}, cell_{f2}^{tree}, h_{f2}^{tree})$. The symbol $\tau$ refers to the embedding of the token of node P. To generate the embedding of descendants of node P, we use the following equations to take $cell_{f1}^{tree}, h_{f1}^{tree}, cell_{f2}^{tree}, h_{f2}^{tree}$ as input and generate $cell_{des}, h_{des}$ as output. 
\begin{equation}\nonumber
[i, j, f_1, f_2, o] = split(W_3 [h_{f1}^{tree}; h_{f2}^{tree}])
\end{equation}
\begin{equation}\nonumber
l_{1} = tanh(i) * sigmoid(j)
\end{equation}
\begin{equation}\nonumber
l_{2} = l_{1} + cell_{f1}^{tree} * sigmoid(f_1)
\end{equation}
\begin{equation}\nonumber
l_{3} = l_{2} + cell_{f2}^{tree} * sigmoid(f_2)
\end{equation}
\begin{equation}\nonumber
cell_{des} = tanh(l_{3})
\end{equation}
\begin{equation}\nonumber
h_{des} = cell_{des} * sigmoid(o)
\end{equation}
In above equations, $W_3 \in \mathbb{R}^{5d \times 2d}$, $i, j, f_1, f_2, o \in \mathbb{R}^{d \times 1}$. The $split$ function splits a tensor $\in \mathbb{R}^{5d \times 1}$ to five tensors $\in \mathbb{R}^{d \times 1}$. With the definition of $cell_{des},\ h_{des}$, the Equation \ref{eq:2d-decode-logit} can be computed. The $cell_{des},\ h_{des}$ for node n are computed based on the encoding of the sub-trees of which the parent is node n. In this subsection, the way to encode the sub-tree (with no child, with 1 child, with 2 or more children) has been introduced. 

\section{Experiment}
Hierarchical Language Model can be applied to any programming language that can be parsed into an abstract syntax tree. 
As the number of projects written in Java programming language is the largest in Github, thus in this experiment, famous Java projects are downloaded from Github to fill into the code corpus (data set) to do experiments. 
The source code of each downloaded project is pre-processed to ensure its quality. 
For each code corpus, the training set accounted for 60\%, the validation set accounted for 15\%, the test set accounted for 25\%. 
In experiments, the validation set is used to prevent over-fitting. 
Every function is parsed into AST and 
every AST will be regarded as a training example or as a test example. 
Each node in an AST is predicted. 
The models in experiments are trained to predict each node in AST correctly. 
The accuracy is the summation of the prediction accuracy of each node in each AST. 
Some sequential models such as RNN or LSTM cannot be directly applied to data in tree structures. 
The tree will be flattened into a sequence, making sequential models applicable. 
\subsection{UNK Setting} 
In natural language processing, the least frequently occurred words are marked as $UNK$. 
In order not to make $UNK$ become the most frequently words, we set about 1 to 3 least frequently appeared words in training set as $UNK$. Thus, $UNK$ can still be rare words not the most frequently words. In test data, the embedding of out-of-vocabulary words is replaced with the embedding of $UNK$ but we do not think $UNK$ is a right word when computing prediction accuracy. 

\subsection{Date Sets} 
In this experiment, three data sets are generated. 
Dataset 1, 2 are inner project code corpus. Dataset 3 is cross project code corpus. 
Table \ref{table-datasets} shows the composition of each data set. 
Dataset 1 consists of Java files in the main module of project \emph{apache commons lang}. The size of Dataset 1 is 2.8MB. 
The \emph{apache maven} is a famous project and we download the source code on its official website (not on GitHub). The size of the project is 4.4MB. 
As observed from open source projects, many files contain a large amount of Java docs, comments or small functions with only one or two statements. Those noisy data should be removed. 
For generating cross-project data sets, we use the following 3 steps to generate high quality data sets which contains long and non-noisy code. 
The first step is to choose two to four projects on Github in random. The second step is to compute a score for each Java file in each project: the total number of nodes in functions dividing the total number of functions results in the score for a Java file. 
Given the threshold of the size of the data set (for example 8MB), the third step is to select the top Java files with highest scores in each project to mix into a data set until the threshold of the size of the data set is reached. 
Dataset 3 are top-scored Java files from projects \emph{Gocd} (5023 stars), \emph{apache-incubator-joshua} (73 stars), \emph{vlc-android-sdk} (723 stars) and \emph{locationtech-geowave} (344 stars). 
In all data sets, functions with less than 100 AST nodes or more than 10000 AST nodes will be removed. 
The evaluation results on Dataset 3 are more convincing because such results reflect the performance of the model on different projects. 
The last column in Table \ref{table-datasets} shows the vocabulary size of each data set. 
\begin{table}[htbp]
\centering
\caption{Datasets}
\label{table-datasets}
\begin{tabular}{|c|c|c|c|}
\hline
          & From Projects                                                                                           & Size  & Vocabulary \\ \hline
Dataset 1 & apache commons lang main                                                                                         & 2.8MB & 1273       \\ \hline
Dataset 2 & apache maven                                                                                         & 4.4MB & 5283       \\ \hline
Dataset 3 & \begin{tabular}[c]{@{}c@{}}gocd \& apache-incubator-joshua \\ \& locationtech-geowave\end{tabular} & 7.53MB & 8030       \\ \hline
\end{tabular}
\end{table}

\subsection{Baselines}  
To evaluate the performance of our model, some baselines are needed to be trained. 
RNN and LSTM are learned to predict next token based on already predicted tokens in the sequence generated by flattening a tree. 
RNN and LSTM are classical models for capturing the patterns in sequential data. These two models are included in the baselines. Compared to RNN, LSTM has more powerful ability to capture the long term dependency in sequential data. 
Every model in baselines need to predict every token in every function in the data set. 

\subsection{Hyper parameters}
We use Adam Algorithm to compute the gradients. 
We train examples one by one instead of grouping examples into
batches, because different ASTs may have different number
of nodes. The vector size for the feature vector of one token is 128. All other parameters can be decided successively. 

\subsection{Termination condition}
We use the strategy of early-stopping as the termination of model training. 
The model would stop instantly if the the prediction accuracy on validation set starts to decrease. 

\subsection{Platform}
The experiments are conducted on desktop computer. The CPU of the computer is i5-8400. The GPU is Geforce RTX 2070. The memory size is 32GB. 
The implementation of the model is based on TensorFlow. 

\subsection{Evaluation}
The metrics to evaluate the performance of different models in this experiment include top-k (top-1, top-3, top-6 and top-10) accuracy and mrr (mean reciprocal rank). 
The top-k accuracy is computed by judging whether the right token is appeared in the first k recommendation of the code completion model. 
When predicting the token of next node, 
if the oracle token is appeared in the $r$th recommendation, then the rank of this recommendation (completion) is r and the reciprocal rank of this recommendation is $\frac{1}{r}$. 
The mrr is computed by averaging the reciprocal rank of the oracle token for each code recommendation (completion). 

Table \ref{table-accuracy-test} shows the top-k accuracy and the mrr evaluated on four data sets. 
HLM refers to the Hierarchical Language Model. 
On all 3 data sets, RNN performs worst. 
On small data sets: Dataset 1 and Dataset 2, HLM averagely achieves 
8.5\% higher top-1 accuracy than LSTM. On large data set: Dataset 3, HLM averagely achieves 16.9\% higher top-1 accuracy than LSTM. 
For top-1 accuracy, different models in experiments have a large degree of discrimination. 
For top-3, top-6 and top-10 accuracy, the distinction between different models is getting smaller and smaller. 
This illustrates that the top-1 prediction accuracy is most convincing. 
From the perspective of top-1 prediction accuracy, HLM performs better than other models. 
\begin{table}[htbp]
\caption{Evaluation Results on Test Set}
\label{table-accuracy-test}
\centering
\begin{tabular}{|l|l|l|l|l|l|l|}
\hline
DS & MD & top1 & top3 & top6 & top10 & mrr \\ \hline
\multirow{3}{*}{1} & RNN & 32.8 & 45.1 & 53.6 & 59.8 & 0.41 \\ \cline{2-7} 
 & LSTM & 46.8 & 60.7 & 67.2 & 71.3 & 0.55 \\ \cline{2-7} 
 & HLM & \textbf{50.3} & 64.7 & 70.7 & 73.8 & 0.58 \\ \cline{2-7} 
\hline
\multirow{3}{*}{2} & RNN & 44.0 & 56.9 & 63.8 & 69.3 & 0.52 \\ \cline{2-7} 
 & LSTM & 50.9 & 64.1 & 69.6 & 72.5 & 0.58 \\ \cline{2-7} 
 & HLM & \textbf{55.7} & 69.5 & 73.4 & 75.2 & 0.63 \\ \cline{2-7} 
 \hline
\multirow{3}{*}{3} & RNN & 34.8 & 52.2 & 56.4 & 59.3 & 0.43 \\ \cline{2-7} 
 & LSTM & 48.6 & 61.3 & 68.6 & 70.4 & 0.56 \\ \cline{2-7} 
 & HLM & \textbf{56.8} & 64.9 & 71.3 & 72.5 & 0.62 \\ \cline{2-7} 
\hline
\end{tabular}
\end{table}

The HLM uses post-order traversal to traverse the AST to encode all sub-trees. This kind of encoding is good at handling unseen data. As the unseen token is often on the leaf node of AST, if we keep abstracting important information from the leaves to the root, the impact of unseen data on leaves are often reduced. 
In the meanwhile, the standard LSTM model treats all tokens equally. When encountering unseen tokens, the standard LSTM model handles the unseen token and the information bias appears. The information bias can be accumulated if there are many unseen data in a long sequence. This problem is called $exponential\ bias$. The ability to reduce the impact of the unseen tokens is the key for the performance improvement. Besides, the multi-dimensional LSTM (actually 2D-LSTM) used in this model can extend the expression space of the model. All these factors contribute to the improvement of the standard LSTM model. 
The hierarchical BiLSTM is used to generate the encoding of a tree. The statement-level BiLSTM has proved to be effective in code classification tasks and performs better than other tree encoding model such as TreeNN or TBCNN.  But it is still necessary to investigate other encoding models such as TreeNN or TBCNN in the task of code completion under the proposed framework. Besides, the proposed encoding model is a hierarchical BiLSTM model which is different from the statement-level BiLSTM. In future work, we will compare the performance of different encoding models under the proposed framework to identify which model is best for code completion. 
In summary, The top-1 prediction accuracy on the test set is representative and HLM performs better in top-1 accuracy than all other models, so we can conclude that HLM outperforms state-of-art models in handling data inconsistency. 

\section{Conclusion}
The Hierarchical Language Model (HLM) is proposed 
to handle the hierarchical structures of the code syntax tree. 
According to experiments, HLM gains at least 7\% improvement of top-1 accuracy compared to LSTM model. 
In future work, we will adopt the methods about eliminating unseen tokens to further improve the performance of this model. 

\begingroup
\let\itshape\upshape
\bibliographystyle{IEEEtran}
\bibliography{reference.bib}
\endgroup

\end{document}